\documentclass[apj, iop]{emulateapj}
\usepackage[]{natbib}
\usepackage{graphicx}

\newcommand{\RV}{V_\mathrm{los}}

\newcommand{\kms}{\mathrm{km\,s}^{-1}}
\newcommand{\teff}{T_\mathrm{eff}}
\newcommand{\vgal}{V_\mathrm{gal}}
\newcommand{\vrad}{V_\mathrm{los}}
\newcommand{\masyr}{\mathrm{mas\,yr}^{-1}}
\newcommand{\kpc}{\mathrm{kpc}}
\newcommand{\myr}{\mathrm{Myr}}
\newcommand{\gyr}{\mathrm{Gyr}}
\newcommand{\magA}{\mathrm{mag}}
\newcommand{\mh}{\hbox{[M/H]}}
\newcommand{\energyunits}{\mathrm{km}^2\,\mathrm{s}^{-2}}
\newcommand{\VPHI}{V_\phi}
\newcommand{\VR}{V_R}
\newcommand{\VZ}{V_Z}
\newcommand{\lamsun}{\Lambda_\odot}

\slugcomment{{\sc Accepted to ApJ:} December 8, 2010}

\shorttitle{The Aquarius Stream}
\shortauthors{Williams et al. }

\begin{document}

\title{The Dawning of the Stream of Aquarius in RAVE}

\author{M. E. K. Williams\altaffilmark{1}, M. Steinmetz\altaffilmark{1}, S. Sharma\altaffilmark{2}, J. Bland-Hawthorn\altaffilmark{3}, R. S. de Jong\altaffilmark{1}, G. M. Seabroke\altaffilmark{4}, A. Helmi\altaffilmark{5}, K. C. Freeman\altaffilmark{6}, J. Binney\altaffilmark{7}, I. Minchev \altaffilmark{1}, O. Bienaym{\'e}\altaffilmark{8}, R. Campbell\altaffilmark{9}, J. P. Fulbright\altaffilmark{10}, B. K. Gibson\altaffilmark{11}, G. F. Gilmore\altaffilmark{12}, E. K. Grebel\altaffilmark{13},  U. Munari\altaffilmark{14}, J. F. Navarro\altaffilmark{15}, Q. A. Parker\altaffilmark{16, 3}, W. Reid\altaffilmark{16}, A. Siebert\altaffilmark{8}, A. Siviero\altaffilmark{14, 1},  F. G. Watson\altaffilmark{3}, R. F. G. Wyse\altaffilmark{10}, T. Zwitter\altaffilmark{17, 18}}
\altaffiltext{1}{Astrophysikalisches Institut Potsdam, An der Sterwarte 16, D-14482 Potsdam, Germany  \email{mailto:mary@aip.de}}
\altaffiltext{2}{School of Physics, University of Sydney, NSW-2006, Australia}
\altaffiltext{3}{Australian Astronomical Observatory, P.O. Box 296, Epping, NSW 1710, Australia}
\altaffiltext{4}{Mullard Space Science Laboratory, University College London, Holmbury St Mary, Dorking, RH5 6NT, UK}
\altaffiltext{5}{Kapteyn Astronomical Institute, University of Groningen, Postbus 800, 9700 AV Groningen, Netherlands}
\altaffiltext{6}{RSAA Australian National University, Mount Stromlo Observatory, Cotter Road, Weston Creek, Canberra, ACT
72611, Australia}
\altaffiltext{7}{Rudolf Pierls Center for Theoretical Physics, University of Oxford, 1 Keble Road, Oxford OX1 3NP, UK}
\altaffiltext{8}{Observatoire astronomique de Strasbourg, Universit\'e de Strasbourg, CNRS, UMR 7550, Strasbourg, France}
\altaffiltext{9}{Western Kentucky University, Bowling Green, Kentucky, USA}
\altaffiltext{10}{Johns Hopkins University, 3400 N Charles Street, Baltimore, MD 21218, USA}
\altaffiltext{11}{Jeremiah Horrocks Institute for Astrophysics \& Super-computing, University of Central Lancashire, Preston, UK}
\altaffiltext{12}{Institute of Astronomy, University of Cambridge, Madingley Road, Cambridge CB3 0HA, UK}
\altaffiltext{13}{Astronomisches Rechen-Institut, Zentrum f\"ur Astronomie der Universit\"at Heidelberg, D-69120 Heidelberg, Germany}
\altaffiltext{14}{INAF Osservatorio Astronomico di Padova, Via dell'Osservatorio 8, Asiago I-36012, Italy}
\altaffiltext{15}{University of Victoria, P.O. Box 3055, Station CSC, Victoria, BC V8W 3P6, Canada}
\altaffiltext{16}{Macquarie University, Sydney, NSW 2109, Australia}
\altaffiltext{17}{University of Ljubljana, Faculty of Mathematics and Physics, Ljubljana,Slovenia}
\altaffiltext{18}{Center of excellence SPACE-SI, Ljubljana, Slovenia}

\begin{abstract}
We identify a new, nearby ($0.5\lesssim d\lesssim 10\,\kpc$) stream in data from the RAdial Velocity Experiment (RAVE). As the majority of stars in the stream lie in the constellation of Aquarius we name it the Aquarius Stream. We identify 15 members of the stream lying between $30^\circ< l <75^\circ$ and $-70^\circ< b <-50^\circ$, with heliocentric line-of-sight velocities $V_\mathrm{los}\sim-200\,\kms$. The members are outliers in the radial velocity distribution, and the overdensity is statistically significant when compared to mock samples created with both the
Besan\c con Galaxy model and newly-developed code Galaxia. The metallicity distribution function and isochrone fit in the $\log g$ - $\teff$ plane suggest the stream consists of a $10\,\gyr$ old population with $\mh\sim-1.0$. We explore relations to other streams and substructures, finding the stream cannot be identified with known structures: it is a new, nearby substructure in the Galaxy's halo. Using a simple dynamical model of a dissolving satellite galaxy we account for the localization of the stream. We find that the stream is dynamically young and therefore likely the debris of a recently disrupted dwarf galaxy or globular cluster. The Aquarius stream is thus a specimen of ongoing hierarchical Galaxy formation, rare for being right in the solar suburb. 
\end{abstract}


\section{Introduction}
\label{sec:Intro}
Under the current paradigm of galaxy formation galaxies build via a hierarchical process and our Galaxy is deemed no exception. Relics of formation are observed as spatial and kinematic substructures in the Galaxy's stellar halo. Recent observations such as those from the Sloan Digital Sky Survey (SDSS) have brought a large increase in the detections of substructures within the outer reaches of the halo (out to $d<80$ kpc). These streams have usually been detected as spatial overdensities from photometry e.g., \citet{Yanny2000, Majewski2003, Belokurov2006, Newberg2009}. Many of these structures have been identified as belonging to the debris of the Sagittarius dwarf spheroidal galaxy (Sgr dSph), which traces the polar orbit of this galaxy as it merges with the Milky Way. Furthermore, after subtracting such prominent substructures \citet{Bell2008} observed a dominant fraction of the halo to deviate from a smooth distribution, consistent with being primarily accretion debris.

Closer to the Sun the spatial coherence of streams and substructures is not so easily discernible and most streams of stars are visible only as velocity structures, such as the \cite{Helmi1999} stream. Indeed, \citet{Helmi2009} has shown that only at distances greater than $\sim10\,\kpc$ do we expect that the structures associated with tidal debris to be observable as spatial overdensities. Therefore, if we wish to identify and study structures within the inner reaches of the halo - where they are most accessible for high resolution follow-up observations - we must search utilizing kinematic data.

Kinematic surveys of the solar neighbourhood are therefore ideal to detect substructures in the nearby regions of the Galaxy's halo. RAVE (RAdial Velocity Experiment) is an ambitious program to conduct a 17,000 square degree survey measuring line-of-sight velocities, stellar parameters, metallicities and abundance ratios of up to 1 million stars \citep{Steinmetz2006}. RAVE utilizes the wide field ($30$ deg$^2$) multi-object spectrograph 6dF instrument on the 1.2-m UK Schmidt Telescope of the Anglo-Australian Observatory (AAO). RAVE's input catalogue for the most part\footnote{Red giants in the direction of rotation were also targeted between $225^\circ < l < 315^\circ$, $5^\circ < |b| < 25^\circ $ with $J-K > 0.5$. This region is not discussed in this paper however.} has only a magnitude selection criterion of $9<I<13$, thus creating a sample with no kinematic biases. The observations are in the Ca-triplet spectral region at 840 nm to 875 nm with an effective resolution of $R=7500$. Starting in April 2003, at the end of 2009 RAVE had collected more than 400,000 spectra. RAVE's radial velocities are accurate to $1.3\,\kms$ when compared to external measurements, while the repeat observations exhibit an accuracy of $2\,\kms$ \citep{Zwitter2008}. These highly accurate radial velocities make RAVE ideal to search for kinematic substructures in an extended region around the sun. Indeed, with RAVE we now move away from studying the \textit{solar neighbourhood} (e.g. \citet{Nordstrom2004}: $d < 0.2\,\textrm{kpc}$) to examining the \textit{solar suburb} ($d < 4\,\kpc$). 

Using RAVE's highly accurate radial velocities, we have discovered a stream that lies mostly within the constellation of Aquarius at a distance of $0.5\lesssim d\lesssim 10\,\kpc$, in the direction $(l,\ b)\sim(55^\circ,\ -60^\circ)$ and at $\RV=-200\,\kms$. The velocity places the stream as part of the Galaxy's halo. As it lies in the direction of the constellation of Aquarius we have named it the Aquarius stream. The detection of this stream is described in Section \ref{sec:Stream}. In Section \ref{sec:Bes} we compare the RAVE data to mock data from the Besan\c con Galaxy mode and the newly-developed galaxy modelling code Galaxia, which offers a number of significant advantages.
Using these models we determine the significance of the detection and constrain its localization. In Section \ref{sec:Pop} we use RAVE's stellar parameters combined with 2MASS ($JHK$) photometry to infer basic properties of the stream population and derive distance estimates. We also use Reduced Proper Motions to obtain another estimate of the distances.
The stream appears to be highly localized on the sky which is interesting considering the apparent proximity of the stream. In Section \ref{sec:Pos} we explore possible connections of the Aquarius stream to other known spatial and kinematic streams, finding that it is not linked to any previously reported structure. In Section \ref{sec:Nat} we investigate possible connections to other (marginal) over-densities in the RAVE dataset, and conclude that the stream is unlikely to be associated with any of them. A simple model of the recent disruption of a satellite in the Galaxy's potential is able to account for the observed localization. The Aquarius stream thus is a new and nearby enigma in the Milky Way's halo.

\begin{figure*}[!tb]
\epsscale{1.0}\plotone{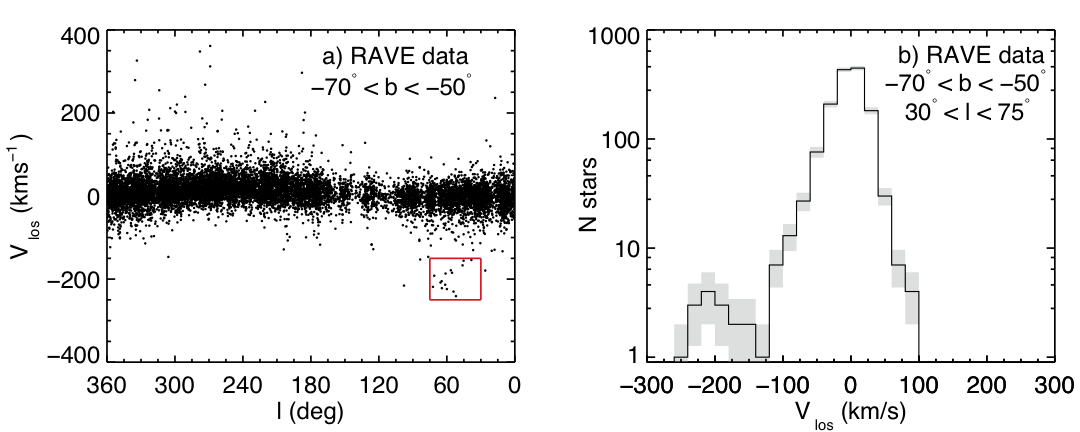}
\caption{(a) $\vrad$ as a function of galactic latitude for RAVE data with $-70<b<-50$, $J>10.3$. The Aquarius Stream is identified as an overdensity of stars with $-250<V_\mathrm{los}<-150\,\kms$, $30^\circ<l<75^\circ$, as delimited by the red box. (b) The histogram of $\vrad$ with the additional constraint $30^\circ<l<75^\circ$ clearly shows the stream as an anomalous feature in the wings of the velocity distribution. The grey shading displays the $\pm 1\sigma$ limits.
\label{f1}}
\end{figure*}

\section{Detection in RAVE}
\label{sec:Stream}
\subsection{The sample}
\label{subsec:Sample}
RAVE measures the velocities of stars that are selected purely on the basis of their photometry, so it is free of kinematic biases. Over most of the sky the probability of a star's selection depends entirely on its apparent magnitude; only in directions towards the Galactic Centre is selection based on colour as well as magnitude (DR1: \citet{Steinmetz2006}, DR2:  \citet{Zwitter2008}). Furthermore, RAVE's radial velocities are accurate to $\le2\,\kms$ so fine substructures are best detected using radial velocities alone: combining them with proper motions and distances mean a significant loss of accuracy. The Aquarius stream was discovered in RAVE data as a structure seen in heliocentric radial velocity vs Galactic latitude/longitude space. When the stream was first noted, it was found to be most clearly defined by faint stars with low gravities, which suggests that the structure is at some distance from the Sun. Removing foreground giants enhances its visibility.

We use the internal release of RAVE from January 2010 that contains 332,747 RVs of 252,790 individual stars. We use only those observations for which the signal-to-noise ratio $SNR>13$ and the Tonry and Davis cross-correlation coefficient $R>5$ to remove potentially erroneous observations. Note that, since not all observations have the more accurate signal-to-noise estimation, $S2N$, we use the $SNR$ value which can underestimate the signal-to-noise (see DR2). For multiple observations of single stars the $\vrad$ were averaged, as were the stellar parameters for those observations that yielded an estimate of these parameters.

The Aquarius stream was found in the Galactic latitude slice $-70^\circ<b<-50^\circ$. As described above, it is also more marked for fainter stars. We therefore introduce an upper brightness limit to enhance the visibility of the stream. As noted in the first and second data release papers, a subset of the RAVE input catalog have $I$ magnitudes from the SuperCOSMOS Sky Survey \citep{Hambly2001}, which show an offset to DENIS $I$ magnitudes. Not all RAVE stars have DENIS $I$ magnitudes either. We therefore turned to 2MASS bands for our magnitude limit, even though this tends to bias against cool stars in our sample, and we potentially miss some candidates. We found that a limit of $J>10.3$ produced the best differentiation of the stream from the background population, removing the brighter, nearby giants.

\subsection{Detected overdensity}
Figure \ref{f1}a shows the structure seen in heliocentric radial velocity, $\RV$, against Galactic longitude, $l$, for the stars with the selection criteria $-70^\circ<b<-50^\circ$, $J>10.3$. A clear structure begins at $\RV\sim-150\,\kms$ at $l=30^\circ$ and extends down to $\RV=-200\,\kms$ at $l=75^\circ$. This overdensity is particularly clear in Figure \ref{f1}b, where we plot the histogram for $\vrad$ in the region $-70^\circ<b<-50^\circ$, $J>10.3$, $30^\circ<l<75^\circ$. The stream can be seen as an excess of stars at negative velocities that is distinct from the general population.

\begin{table*}[!t]
\begin{tiny}
\begin{tabular}{llllllllllll}
\hline
ID&RA&DEC&Obsdate&$\RV$&e$\RV$&$V_\mathrm{gal}$&$\mu_\alpha$&$e\mu_\alpha$&$\mu_\delta$&$e\mu_\delta$&SNR\\
 &          &        &                & $\kms$&$\kms$&$\kms$&$\masyr$&$\masyr$& $\masyr$&$\masyr$&            \\
\hline
J221821.2-183424 & $22^\mathrm{h}18^\mathrm{m}21.20^\mathrm{s}$ & $-18^\circ2064.5"$ & 20060602 & -154.1 &  1.1 &  -70.7 &  -2.9 &   5.0& -1.3 &   5.0& 32.0\\
C2222241-094912 & $22^\mathrm{h}22^\mathrm{m}24.10^\mathrm{s}$ & $-09^\circ2952.6"$ & 20030617&-241.0& 2.6&-127.6& 32.9&  4.0&-55.2&  4.0& 18.8\\
C2225316-145437 & $22^\mathrm{h}25^\mathrm{m}31.70^\mathrm{s}$ & $-14^\circ3277.9"$ & 20040628 & -155.7 &  0.7 &  -60.8 &  -2.5 &   2.8&-15.3 &   2.7& 33.3\\
C2233207-090021 & $22^\mathrm{h}33^\mathrm{m}20.80^\mathrm{s}$ & $-09^\circ  21.4"$ & 20030617&-184.8& 4.3& -71.6&  5.8&  2.9& -7.5&  2.9& 15.0\\
C2234420-082649 & $22^\mathrm{h}34^\mathrm{m}42.00^\mathrm{s}$ & $-08^\circ1609.5"$ & 20030618 & -177.1 &  1.4 &  -62.4 &  -1.0 &   2.1&-25.2 &   2.1& 25.3\\
C2234420-082649 & $22^\mathrm{h}34^\mathrm{m}42.00^\mathrm{s}$ & $-08^\circ1609.5"$ & 20050914 & -180.4 &  1.1 &  -65.7 &  -1.0 &   2.1&-25.2 &   2.1& 22.1\\
J223504.3-152834 & $22^\mathrm{h}35^\mathrm{m}04.40^\mathrm{s}$ & $-15^\circ1714.9"$ & 20060624 & -166.9 &  1.3 &  -76.5 &   3.4 &   2.1&-14.7 &   2.2& 18.2\\
C2238028-051612 & $22^\mathrm{h}38^\mathrm{m}02.80^\mathrm{s}$ & $-05^\circ 972.9"$ & 20050807 & -213.6 &  1.6 &  -89.4 &  -2.3 &   4.0& -7.4 &   4.0& 14.8\\
J223811.4-104126 & $22^\mathrm{h}38^\mathrm{m}11.50^\mathrm{s}$ & $-10^\circ2487.3"$ & 20060804 & -230.1 &  1.9 & -123.9 &  28.5 &   2.7& -2.0 &   2.7& 22.3\\
C2242408-024953 & $22^\mathrm{h}42^\mathrm{m}40.80^\mathrm{s}$ & $-02^\circ2993.9"$ & 20050909 & -208.3 &  1.5 &  -77.8 &   1.1 &   4.0& -3.7 &   4.0& 14.8\\
C2246264-043107 & $22^\mathrm{h}46^\mathrm{m}26.50^\mathrm{s}$ & $-04^\circ1867.2"$ & 20050807 & -205.0 &  1.7 &  -81.0 & -10.6 &   2.5&-19.3 &   2.5& 20.5\\
C2306265-085103 & $23^\mathrm{h}06^\mathrm{m}26.60^\mathrm{s}$ & $-08^\circ3063.8"$ & 20030907&-221.8& 1.7&-118.7& 15.9&  2.2&-12.8&  2.2& 25.3\\
C2309161-120812 & $23^\mathrm{h}09^\mathrm{m}16.10^\mathrm{s}$ & $-12^\circ 492.0"$ & 20040627 & -224.1 &  2.1 & -133.1 & -25.3 &   2.1&-99.5 &   2.1& 14.6\\
C2322499-135351 & $23^\mathrm{h}22^\mathrm{m}50.00^\mathrm{s}$ & $-13^\circ3231.5"$ & 20040627 & -186.6 &  1.3 & -106.8 &  -2.8 &   2.7& -8.8 &   2.7& 14.7\\
J232320.8-080925 & $23^\mathrm{h}23^\mathrm{m}20.90^\mathrm{s}$ & $-08^\circ 566.1"$ & 20060915 & -191.9 &  1.2 &  -93.0 &  31.1 &   2.0&-58.2 &   2.1& 20.2\\
J232619.4-080808 & $23^\mathrm{h}26^\mathrm{m}19.50^\mathrm{s}$ & $-08^\circ 488.7"$ & 20060915 & -218.7 &  0.7 & -120.9 &  12.3 &   4.0&-24.7 &   4.0& 26.1\\
\end{tabular}
\end{tiny}
\caption{The Aquarius Stream candidates selected from the RAVE data and their parameters. The proper motions are from PPMXL.}
\label{tab1}
\end{table*}

We establish limits of $-250<\RV<-150\,\kms$, $30^\circ<l<75^\circ$, $J>10.3$ to choose 15 candidates of the Aquarius stream, which are outlined by the red box in Figure \ref{f1} and listed in Table \ref{tab1}. Many stream candidates lack stellar parameter estimates, since they were observed early on by RAVE (DR1 does not include such estimates; see the data release papers for details). The average SNR is 20 for the stream candidates and 1 star (C2234420-082649) has a repeat observation, which is listed to show the consistency of the $\vrad$ results. As a double-check, the template fits to each of the spectra were eye-balled as were the zero-point fits (using sky radial velocities) for the fields the stars were observed in. No abnormalities were detected. 

The RAVE internal release includes PPMX proper motions \citep{Roeser2008}. However, for our stream candidates we use in the following analysis PPMXL proper motions \citep{Roeser2010}), where the average proper motion error for the stream stars is reduced from $e_{\mu}=6.8\,\masyr$ in PPMX to $e_{\mu}=4.3\,\masyr$ in PPMXL. These proper motions are also listed in Table \ref{tab1}.

The average heliocentric radial velocity of the stream is $\RV=-199\pm27\,\kms$ and its Galactocentric radial velocity, i.e. the line-of-sight velocity in the Galactic rest frame (see equation 10-8 of \citet{BinneyTremaine}), is $\vgal=-93\pm25\,\kms$. When compared to $\RV=-120\pm100\,\kms,\ \vgal=0\pm100\,\kms$ for the halo and $\RV=-30\pm45\,\kms,\ \vgal=90\pm45\,\kms$ for the thick disk at (l,\ b)=($55^\circ,\ -60^\circ$), this velocity indicates that the group to be a halo feature. However, it still has quite a large velocity even for the halo.

\section{Model comparisons}
\label{sec:Bes}
\subsection{Besan\c con and Galaxia models}
\label{subsec:models}
To establish the statistical significance of the Aquarius overdensity we compare the RAVE sample to mock samples created using the Besan\c con Galaxy model \citep{Robin2003} and the newly developed galaxy modeling code Galaxia \citep{Sharma2010}. Galaxia is based on the Besan\c con Galaxy model, but with several improvements. The first is a continuous distribution created across the sky instead of discrete sample points. Second is the ability to create samples over an angular area of arbitrary size. Third, it utilizes Padova \citep{Girardi2002} isochrones which offer support for multiple photometric bands. Fourth, Galaxia offers greater flexibility with dust modelling. Once a data set without extinction has been created, multiple samples with different reddening normalization and modelling can be easily generated.  Finally, with Galaxia multiple independent random samples can be generated, which is crucial for doing a proper statistical analysis. Due to the above mentioned advantages we chose Galaxia as our preferred model to create mock samples.

\begin{figure*}[!tb]
\epsscale{1.0}\plotone{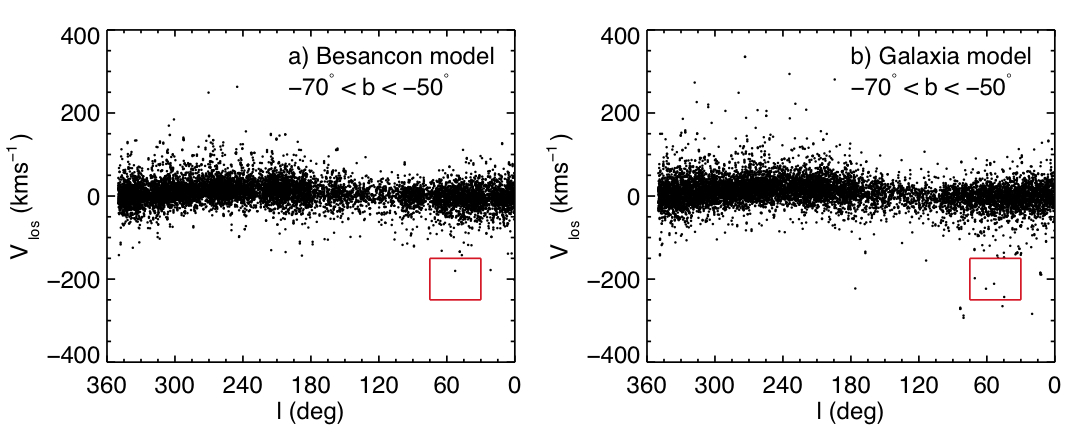}
\caption{(a) The mock Besan\c con sample and (b) a mock Galaxia sample for $-70<b<-50$, $J>10.3$. As in Figure 1, the Aquarius stream region is delimited by the red box, with both mock samples displaying a a paucity of stars in this region. A reddening rate of $E(B-V)$=0.23 mag/kpc is used for both the model samples displayed (see Section \ref{sec:Bes} for details).
\label{f2}}
\end{figure*}

 \begin{table*}[!ht]
\begin{footnotesize}
\begin{tabular}{lllll}
\hline
Model&Solar position&Solar motion&$V_c$&$E(B-V)$ rate\\
 &$(x,\ y,\ z)$ kpc&$(U,\ V,\ W)\,\kms$&$\kms$&mag/kpc\\
 \hline
Besan\c con&$(-8.5,\ 0.0,\ 0.015)$&$(10.30,\ 6.30,\ 5.90)$&226.40& 0.23\\
Galaxia&$(-8.0,\ 0.0,\ 0.015)$&$(11.10,\ 12.24,\ 7.25)$&226.84&0.23, 0.53, Schlegel\\
\end{tabular}
\caption{Parameters for the Galaxia and Besan\c con models used for comparison with the RAVE sample. }
\label{tab2}
\end{footnotesize}
\end{table*}

Table \ref{tab2} lists the basic parameters for each of the two models. For the dust modelling, we chose the default value for the Besan\c con model, where the dust is modelled by an Einasto disk with a normalization of $A_V=0.7\,\magA/\kpc$. This is reasonable for the high latitudes that we simulate. Assuming a $R_V$=3.1, this corresponds to a reddening rate of $E(B-V)=0.23\,\magA/\kpc$. No additional dust clouds were added. For the Galaxia model, we present results with the dust modelled by an exponential disk, with the reddening rate in the solar neighbourhood normalized to 0.23 and $0.53\,\magA/\kpc$, where the latter is taken from \cite{BinneyMerrifield}. Also, we present results for a model where the reddening at infinity is matched to that of the value in Schlegel maps. To convert $E(B-V)$ to extinction in different photometric bands we used the conversion factors in Table 6 of \citet{Schlegel1998}.

\subsection{Mock sample generation}
\label{subsec:mock}
The mock samples were created from Galaxia and Besan\c con using analogous methodology. Firstly, to create the Besan\c con sample we queried $\Delta l\times \Delta b=50^\circ\times20^\circ$ regions using the online query form imposing the $I$-band magnitude limits of RAVE of $9<I<13$, making no biases in spectral type. A distance limit of $d=20\,\kpc$ is imposed as most RAVE stars (with the exception of a few notable LMC stars - see \citet{Munari2009}) should be within 15 kpc \citep{Breddels2010}. Grid-steps of $10^\circ$ in $l$ and $5^\circ$ in $b$ were used in the query.

To generate samples from Galaxia we simply generated a full catalog over the area specified by $0<l<360$, $-90<b<0$ and $9<I<13$ and then extracted the required samples from it after correcting for extinction. Since Galaxia allows oversampling, the initial catalog was generated with an oversampling factor of 10, so that later on 10 independent random realizations could be created.

Using Monte Carlo techniques each model was then resampled first to create a uniform distribution in $I$-magnitude and then resampled again to exactly mimic the shape of the DENIS $I$-band distribution in a $\Delta l\times \Delta b=50^\circ\times20^\circ$ region. This ensures that the distance distribution will be similar to the RAVE sample. Each generated sample is then further reduced to those stars with $J> 10.3$ to mimic our sample selection in Section \ref{sec:Stream}. Finally, the number of stars in the mock sample is normalized to that of the RAVE sample in sub-regions of $\Delta l\times \Delta b=25^\circ\times10^\circ$, where this division into sub-regions was required to better suit the curved boundary of the RAVE survey area. For the Besan\c con sample, the $l$ and $b$ co-ordinates were smeared out to remove the discretization by adding a uniform randomization of the same extent (since the Galaxia sample was already smoothly distributed no such procedure was required). Also, for Galaxia ten mock
data samples were created for each dust modelling scenario, enabling a better handle on the statistical significance of the Aquarius stream. Finally, to simulate the RAVE radial velocity measurement errors a scatter of $\sigma=2\,\kms$ was added to the models' radial velocities.

\begin{figure*}[!tb]
\epsscale{1.0}\plotone{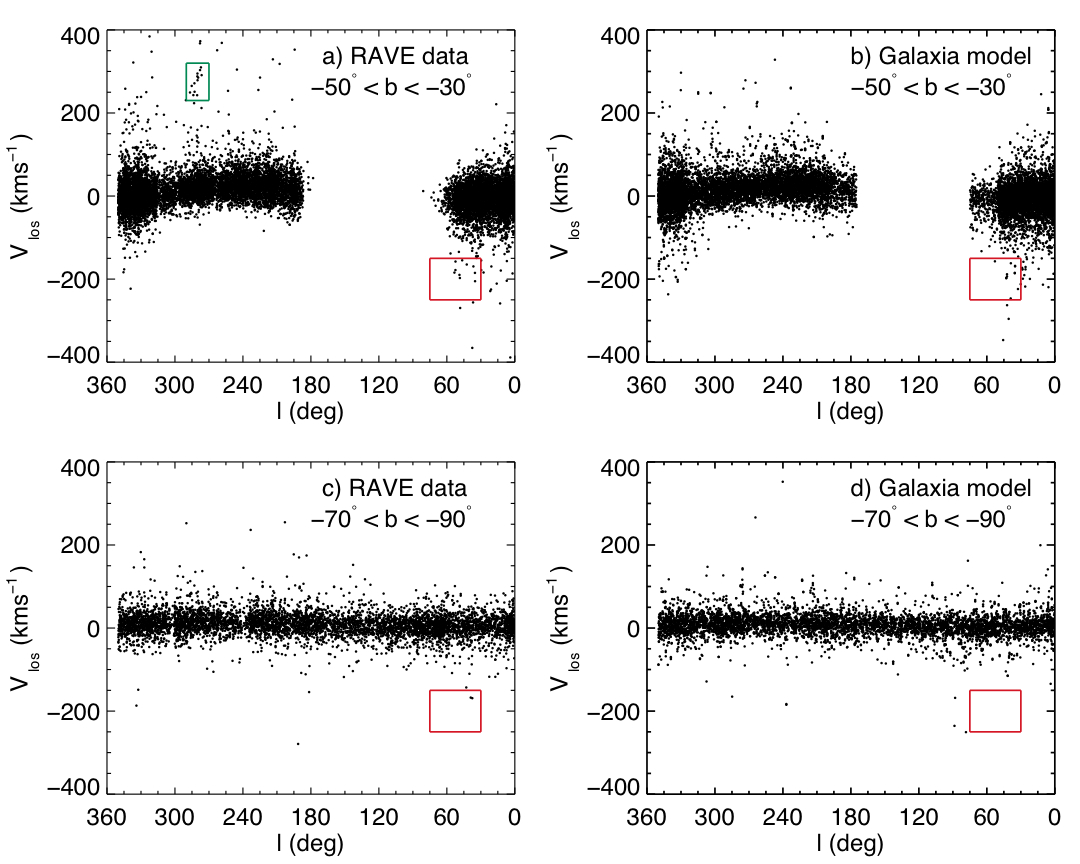}
\caption{As in Figure \ref{f1} but for the latitude ranges $-50<b<-30$ (a, b: top) and $-90<b<-70$ (c, d: bottom) using the Galaxia model with \citet{Schlegel1998} dust mapping for comparison. LMC stars can be seen at $270^\circ<l<290^\circ$, $230<\RV<310\,\kms$ in the RAVE data in the top panel and are outlined by the green box. The placement of the Aquarius stream from Figure \ref{f1} is outlined by the red box. Other than the LMC structures are not easily discernible.
\label{f3}}
\end{figure*}

\subsection{Statistical significance of Aquarius}
\label{subsec:statsig}
Figure \ref{f2} shows the Besan\c con and one of the Galaxia samples (with $E(B-V)=0.23\,\magA/\kpc$) for the same area of the sky as in Figure \ref{f1}a. We see that both models do a fair job of reproducing the gross features of the data. A detailed analysis of comparing both the Besan\c con and Galaxia models to RAVE will be presented in an upcoming papers by A. Ritter. In this analysis it is sufficient to note firstly that the Galaxia model produces a better representation of the density of halo stars (i.e., those stars with larger $\vrad$) than the Besan\c con model. Moreover, the Galaxia model better reproduces the $\RV$ distribution as a function of Galactic latitudes than Besan\c con: for bins of $25^\circ$ in Galactic latitude, on average Galaxia agrees with the data to within $2\,\kms$ and $3\,\kms$ for mean and dispersion in $\RV$ respectively, compared to $3\,\kms$ and $4\,\kms$ for Besan\c con.

To compare the generated samples to the RAVE sample, we establish cells of size $\Delta l\times \Delta \RV$ and for each cell compare the number of stars from RAVE and the mock samples. For each sample in the $i$-th cell there are $N^{\mathrm{Model}} _i$ stars and we estimate the standard deviation by $\sigma_i=\sqrt{N^{\mathrm{Model}}_i}$. We consider an overdensity significant if 
\begin{equation}
N^\mathrm{RAVE} _i - N^\mathrm{Model} _i > 4\sigma_i
\end{equation}
where $N^\mathrm{RAVE} _i$ are the number of RAVE stars in the $i$-th cell and $N^\mathrm{Model}_i$ is either $N^\mathrm{Bes}$ or $N^{\mathrm{Gal},q}_i$, where $q=1..10$ in the latter signifies the sample number from Galaxia. Following a procedure similar to \citet{Helmi1999}, we identify overdense regions in the Galactic latitude slice $-50<b<-70$ by varying the cell sizes with longitude slices ranging from $\Delta l=25,\ 35,\ 50,\ 70$ and radial velocity bins ranging from $\Delta \RV= 20,\ 25,\ 35...100$. We then evaluate the percentage of the various cell sizes which identify the region around $30^\circ<l<75^\circ$, $-250<\RV<-150\,\kms$ as having a $4\sigma$ deviation. As we have 10 samples for Galaxia, we take the average over all the samples, obtaining a mean and standard deviation for this value.

The following results are found: using the Besan\c con model, $96\%$ of the different cell sizes identify that the number of stars in the data are $4\sigma$ overdense around Aquarius compared to the model. For Galaxia using $E(B-V)=0.23\,\magA/\kpc$, we find that $80\pm15\%$ of cell sizes give Aquarius as a $4\sigma$ deviation, while $E(B-V)=0.53\,\magA/\kpc$ gives $75\pm15\%$ and the Schlegel results yield $80\pm18\%$. How the dust is modelled at these high Galactic latitudes therefore has little impact on the results. In general we can conclude that the models robustly show that there is a statistically significant concentration of stars at the Aquarius stream's location. Indeed, for some cell sizes and models the overdensity can be as high as $11\sigma$. This confirms what can be seen by eye: the stream as an overdensity in the outlying regions of the velocity distribution. 

\subsection{Localization of the stream}
\label{subsec:Loc}

In addition to identifying the statistical significance of the Aquarius stream, we also used the models to search for additional members of the stream and possible related substructures. We compared the RAVE data and mock Galaxia samples for surrounding latitude cuts of $-50^\circ<b<-30^\circ$ and $-90^\circ<b<-70^\circ$, where once again we consider only those stars with $J>10.3$. Figure \ref{f3} displays the data in these two latitude ranges compared to Galaxia models using the Schlegel dust model. We repeated the analysis of Section \ref{subsec:statsig}, looking for overdensities in the RAVE data compared to the Galaxia models, varying the cell size and identifying regions with repeated $4\sigma$ signals. 

\begin{figure*}[!tb]
\epsscale{1.0}\plotone{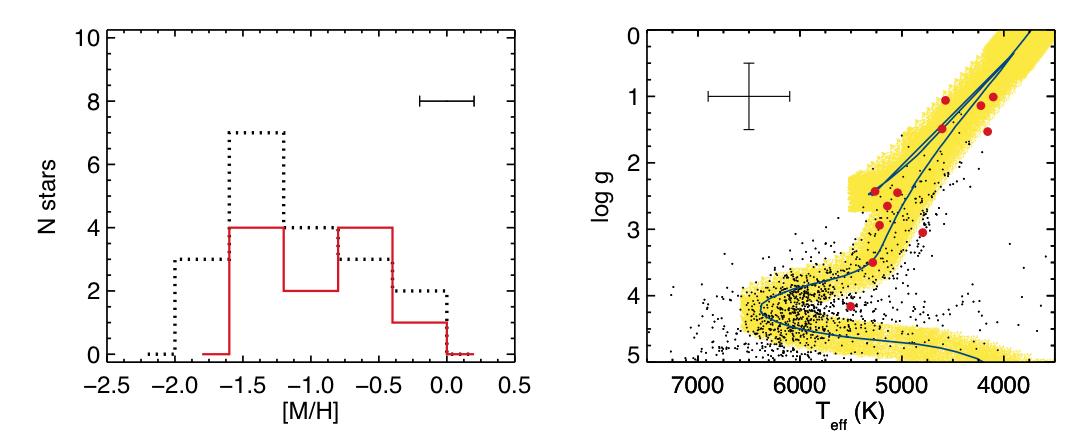}
\caption{\textit{Left:} MDF for the Aquarius stream members (red line), whose typical metallicity uncertainty is $\Delta(\mh)\sim0.2$ dex ($\pm1\sigma$ shown). The MDF of other halo stars with $-70^\circ<b<-50^\circ$, $J>10.3$, $|\vrad |>200\,\kms$ is shown for comparison (dotted line). \textit{Right:} $\teff$ - $\log g$ plane for RAVE stars in the region $-70^\circ<b<-50^\circ$, $30^\circ<l<75^\circ$, $J>10.3$. Stream candidates are highlighted as solid red points and a Padova isochrone with $10\,\gyr$, $\mh=-1$ overplotted. The yellow region indicates $1\sigma$ in both $\teff$ and $\log g$ from this isochrone.
\label{f4}}
\end{figure*}

For the  $-50^\circ<b<-30^\circ$ sample, the region around $270^\circ<l<290^\circ$, $230<\RV<310\,\kms$ is consistently identified for all the various dust models as significantly overdense: on average $95\%$ of cell sizes identify this region as containing a $4\sigma$ signal. These stars are associated with the Large Magellanic Cloud (LMC) and it is reassuring that our technique detects it.

For both latitude ranges, $-50^\circ<b<-30^\circ$ and $-90^\circ<b<-70^\circ$, there are no detections of statistically significant overdensities in the vicinity of the Aquarius stream's velocity and longitude range.  Also, for the $-90^\circ<b<-70^\circ$ sample no particular region has consistent $4\sigma$ deviations when compared to the Galaxia models. The region around $-50^\circ<b<-30^\circ$, $320^\circ<l<350^\circ$, $150<\RV<300\,\kms$ is detected in $\sim50\%$ of the trials as being overdense for this latitude cut, irrespective of dust modelling. A similar detection is also found for the region $-70<b<-50$, $260^\circ<l<340^\circ$, $100<\RV<300\,\kms$, in the same latitude range as the Aquarius stream. These detections are not as significant as Aquarius and are in a different region of the sky.

In general, we find that there are no stars clearly associated to the Aquarius stream in adjacent latitude cuts; no overdensities were detected in the vicinity of the stream's velocity and longitude. This may be caused in part by the survey boundary, but the sharp localization of the stream is nevertheless intriguing. In Section \ref{sec:Nat} we further investigate the localization of the Aquarius stream, examining its possible relation to the two marginal overdensities detected above. 

\section{Population properties of the Aquarius stream}
\label{sec:Pop}

\subsection{Metallicity and $\log g$-$\teff$ plane }
\label{subsec:mdfcmd}

RAVE gives estimates of stellar parameters from the spectra which we can use to establish the basic properties of the population of the Aquarius stream. Conservatively the stellar parameters are accurate to $\sim0.2$ dex in $\mh$,  $400\ \mathrm{K}$ in $T_\mathrm{eff}$ and $0.5$ dex in $\log g$ when compared to external measurements, though internally the errors are significantly smaller \citep{Zwitter2008}. For 13 of the 16 stream candidates we have estimates of stellar parameters.

Figure \ref{f4} shows the metallicity distribution function (MDF) and the $\log g$-$\teff$ plane of these stars where we compare the latter to the background population. The conservative estimates of the errors are also shown. Note that we do not apply the metallicity calibration of Equation 20 in \citet{Zwitter2008} as this calibration does not extend down to halo metallicities $\mh<-1.5$. Therefore, the derived MDF is best seen relative to background halo stars. We plot the MDF for stars selected $-70^\circ<b<-50^\circ$, $l\le30^\circ$ or $l\ge75^\circ$, $J>10.3$, $|\vrad |>200\,\kms$. The stream's MDF peaks at a slightly higher metallicity than these background halo stars with a slightly tighter distribution: the stream has an average $\mh=-1\pm 0.4$ compared to $\mh=-1.1\pm 0.6$ for the background. Both distributions show metallicities with are rather high for the halo. The RAVE 3rd data release, Siebert (2010, in preparation) show that these RAVE stellar parameters tend to overestimate the metallicity of stars with low SNR, and effect that would be on the order of 0.1 dex for the stream stars. This data release will present improved stellar parameter from a modified pipeline, as well as a new metallicity calibration from an extended metallicity range. Hence, these results should a better handle on the stream's MDF. Clearly, however, follow-up high-resolution spectroscopy is required to derive accurate abundances to better understand the group's chemical abundance properties. Nevertheless, from the initial RAVE metallicities, we can conclude that the stream's MDF is consistent with background halo stars. 

Using Padova isochrones \citep{Girardi2002} we find the best fitting isochrone to be that of $10\,\gyr$, $\mh=-1$ which is over-plotted in Figure \ref{f4} (\textit{right}), as well as a highlighted region showing the $\pm 1\sigma$ bounds in $(\log g,\ \teff)$ from this curve. Most of the stream stars fall within this region, though clearly the isochrone fit is preliminary given the size of the stellar parameter errors. From both the MDF and the isochrone fit, however, there is a general indication that the Aquarius stream is metal-poor and old. 

\subsection{Isochrone derived distances}
\label{subsec:iso}
We use the isochrone fit from above to derive distances to the candidate stars, using the $J$-band magnitude. To derive $M_J$ from the isochrone we find the nearest point along the isochrone to the actual data point by minimizing the distance in $\log g$ and $\teff$ between them, normalized by the standard error in each. Extinction is of the order of $A_J=0.04 \pm 0.01$, and is calculated iteratively from the distances using \citet{Schlegel1998} dust maps and assuming a Galactic dust distribution as in \citet{Beers2000}. Errors are calculated via Monte Carlo, generating 100 points around the data point with $\sigma_{T_\mathrm{eff}}=400\ \mathrm{K}$ and $\sigma_{\log g}=0.5$ dex, propagating through to a distribution of distances, from which a standard deviation is derived.

\begin{table*}[!t]
\begin{tiny}
\begin{tabular}{lllllllllllll}
\hline
ID&Obsdate&$J$&$K$&$T_\mathrm{eff}$&$\mh$&$\log g$&$d_\mathrm{I}$&$d_\mathrm{R}$&$d_\mathrm{B}$&$d_\mathrm{Z}$&$d_\mathrm{BB}$\\
 &       &             &                  &K &                        &           &kpc&kpc&kpc&kpc&kpc\\
 \hline
J221821.2-183424 & 20060602 & 10.34& 9.68&4572 & -1.54 & 1.06 & $ 5.6\pm 2.1$ & $ 2.4\pm 1.0$&$ 9.6\pm 2.1$&$ 6.7\pm 2.9$&$ 2.8\pm 0.5$\\
C2222241-094912 & 20030617&10.64& 9.79&-&-&-&- & $16.3\pm11.9$&$ -$&$ -$&$ -$\\
C2225316-145437 & 20040628 & 10.34& 9.57&4104 & -1.29 & 1.01 & $ 7.3\pm 2.7$ & $ 3.4\pm 1.5$&$10.3\pm 2.4$&$ 7.9\pm 3.1$&$ 1.4\pm 0.1$\\
C2233207-090021 & 20030617&11.66&11.28&-&-&-&- & $ 0.8\pm 0.3$&$ -$&$ -$&$ -$\\
C2234420-082649 & 20050914 & 10.67&10.13&5263 & -2.02 & 2.43 & $ 1.9\pm 0.7$ & $ 1.9\pm 0.8$&$ -$&$ 3.1\pm 2.8$&$ -$\\
J223504.3-152834 & 20060624 & 10.36& 9.65&4795 & -0.33 & 3.05 & $ 1.0\pm 0.5$ & $ 2.5\pm 1.1$&$ -$&$ 1.3\pm 1.0$&$ -$\\
C2238028-051612 & 20050807 & 11.53&10.74&4606 & -0.86 & 1.49 & $ 7.1\pm 2.7$ & $ 6.7\pm 5.4$&$ -$&$ -$&$ -$\\
J223811.4-104126 & 20060804 & 10.42& 9.90&5502 & -0.78 & 4.16 & $ 0.4\pm 0.1$ & $ 0.5\pm 0.2$&$ 0.3\pm 0.1$&$ 0.5\pm 0.9$&$ 0.4\pm 0.3$\\
C2242408-024953 & 20050909 & 11.63&10.82&4159 & -0.75 & 1.53 & $ 9.4\pm 3.4$ & $13.4\pm 8.7$&$ -$&$ 8.5\pm 5.6$&$ -$\\
C2246264-043107 & 20050807 & 11.26&10.72&5142 & -1.22 & 2.65 & $ 1.8\pm 0.9$ & $ 1.8\pm 0.7$&$ 3.9\pm 2.1$&$ 3.2\pm 3.0$&$ 2.0\pm 0.8$\\
C2306265-085103 & 20030907&10.31& 9.47&-&-&-&- & $ 3.3\pm 1.5$&$ -$&$ -$&$ -$\\
C2309161-120812 & 20040627 & 10.68& 9.97&5219 & -0.66 & 2.94 & $ 1.0\pm 0.6$ & $ 5.5\pm 2.9$&$ -$&$ 1.5\pm 1.4$&$ -$\\
C2322499-135351 & 20040627 & 10.82&10.28&5043 & -0.64 & 2.45 & $ 1.9\pm 0.9$ & $ 2.1\pm 0.8$&$ -$&$ 2.3\pm 1.6$&$ -$\\
J232320.8-080925 & 20060915 & 10.96&10.47&5286 & -1.10 & 3.50 & $ 0.7\pm 0.4$ & $ 0.8\pm 0.3$&$ 1.0\pm 0.5$&$ 1.8\pm 2.2$&$ 1.1\pm 0.7$\\
J232619.4-080808 & 20060915 & 10.51& 9.76&4225 & -1.22 & 1.14 & $ 6.7\pm 2.7$ & $ 5.6\pm 2.9$&$10.6\pm 2.8$&$ 7.1\pm 3.0$&$ 3.5\pm 0.6$\\
\end{tabular}
\caption{The Aquarius Stream candidates selected from the RAVE data and their parameters. The distances $d_{I}$ are derived from the isochrone in Figure \ref{f4} while $d_\mathrm{R}$ are derived in Section \ref{sec:RPM}. The extra distances are $d_\mathrm{B}$ \citep{Breddels2010},  $d_\mathrm{Z}$ \citep{Zwitter2010} and $d_\mathrm{BB}$ \citep{Burnett2010}. }
\label{tab3}
\end{tiny}
\end{table*}

The distances are listed in Table \ref{tab3} as $d_{I}$, where the distances range from 0.4 to 9.4 kpc (distance moduli: $m-M=8.3$ to $m-M=14.9$), with a mean distance of $d_\mathrm{av}=3.8\pm3.2\,\kpc$ ($m-M=12.9$). There is a hint of a bimodal population of closer (sub- and red clump giants) and farther stars (tip of the giant branch). However, given the uncertainties in the stellar parameters these distances are uncertain and the reality of this bimodality is therefore debatable; in the next section we develop another distance estimate which has a smoother distribution function. 

The large distance range raises the question whether the Aquarius stream is a distinct entity or comprised of multiple structures. The high-resolution abundances mentioned above would help answer this question by ascertaining if the group has any distinctive chemical signatures compared to other halo stars. Occam's razor would weigh against two structures forming this localized stream however. Further, in Section \ref{subsec:sat} we develop a model for the Aquarius stream under the assumption of a single satellite dissolving in the Galaxy's potential. The model predicts that the stream is spread in $XYZ$ away from the sun, with distance $d_\mathrm{model}=3.2\pm0.8\,\kpc$ in the direction $30^\circ<l<75^\circ, -70^\circ<b<-50^\circ$. The distance range derived above therefore probably reflects more on the distance errors than the real distribution for the stream. We assume that the Aquarius stream is a single, distinct object.

The isochrone from Figure \ref{f4} has a $I$-band turn-off of $M_I=3.5$. Hence, for the distance moduli above we could expect turn-off stars in the range $I=11.8$ - $18.5$. The lower magnitude falls within the RAVE magnitude limits ($9<I<13$). However, RAVE's unbiased selection criteria mean that the thin disk dominates dwarf/turn-off stars, even at these higher magnitudes. Our sample of halo dwarfs is therefore too small to detect the turn-off, and we only see giant stars in our Aquarius stream sample from RAVE.

\begin{figure}[!tb]
\epsscale{1.0}\plotone{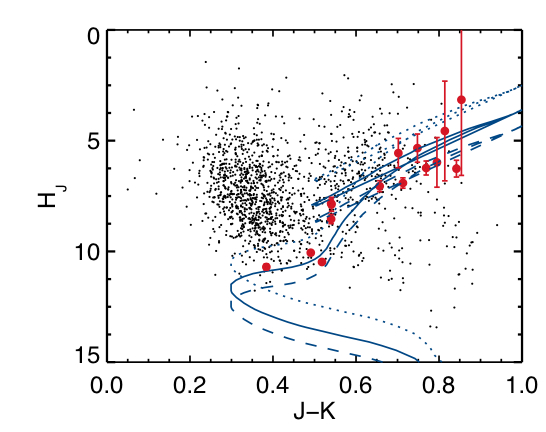}
\caption{Reduced proper motion diagram for the background RAVE stars (black points) and the Aquarius stream stars (red points). The isochrone from Figure \ref{f4} is plotted with a tangential velocity of $v_T=250\,\kms$ (solid line), $v_T=150\,\kms$ (dotted line) and $v_T=350\,\kms$ (dashed line). The coherency of the group is clear also in this diagram.
\label{f5}}
\end{figure}

\subsection{Reduced Proper Motion Diagram}
\label{sec:RPM}
The coherence of the group selection is shown by the reduced proper motion diagram (RPMD), which plots the reduced proper motion (RPM) against color. Described in detail in \citet{Seabroke2008}, the RPMD essentially creates a HR diagram from the proper motions, where the absolute magnitude is smeared by the variation in the tangential speed of the stars. Halo stars have a large dispersion in tangental velocity and so this smearing is large. In contrast, for a small, nearby section of a stream the transverse velocity spread is small and we effectively recover magnitudes for the stars. The RPM is given by
\begin{equation}
H_J=J+5\log \mu +5 = M_J +5 \log v_\mathrm{T}-3.379
\label{Eqn1}
\end{equation}
where $J$ and $M_J$ are the apparent and absolute magnitudes respectively, $\mu$ is the proper motion in arcsec $\mathrm{yr}^{-1}$, $v_T$ is the tangential velocity in $\kms$. Here we have again used 2MASS colors. Thus, from the observables $J$ and $\mu$ we can establish something about the more fundamental parameters $M_J$ and $v_\mathrm{T}$ without requiring a either distance or a radial velocity. Figure \ref{f5} gives the RPMD for the stars in our magnitude and latitude selected sample with the Aquarius stream candidates over-plotted, where for the latter the more accurate PPMXL proper motions were used. Note that for the distances of these stars the reddening will also be of the order of $E(J-K)\sim0.02$, which does not effect the plot significantly and is neglected. The isochrone from Figure \ref{f4} is over-plotted, where we find that a large tangential velocity of $v_T=150\,\kms$ to $350\,\kms$ is required to shift the isochrone to a reasonable fit, which compare to $v_T\sim230\pm100\,\kms$ for the halo for (l,\ b)=($55^\circ,\ -60^\circ$). Once again, the group is consistent with a halo stream.

We will see in Section \ref{sec:Nat} that the tangential velocity for the stream is indeed within a relatively narrow range as shown in Figure \ref{f5}. A few of the redder, more-distant giants deviate from the rest of the group but they also have larger errors in their proper motions, which translates into larger RPM errors as shown: they are within $2\sigma$ of the group fit. The consistency of the fit for the bluer (nearer) stars supports their inclusion in the candidate list, though the range in values for the RPM of halo stars means that we cannot exclude contamination from the halo. Indeed, from Figure \ref{f1} it is clear that we expect a few of the stars to be non-members. We therefore take the consistency of the RPMD to be a good indication of the consistency of the Aquarius member selection but not absolute proof of membership.

\subsection{RPM derived distances}
\label{sec:rpmdist}
If we accept the group's tangential velocity of $v_T=250\,\kms$, we can use the RPM to establish a second estimate of the distance to the stars. From Equation \ref{Eqn1} we have the distance modulus 
\begin{equation}
J-M_J=J-H_J+5 \log v_\mathrm{T}-3.379
\label{Eqn2}
\end{equation}
From this we have distance moduli ranging from 8.5 to 16 for the group members. The corresponding distances are listed in Table \ref{tab3} as $d_\mathrm{R}$, with the errors calculated using the upper and lower tangential velocity bounds as well as the proper motion errors in $H_J$. These distances differ somewhat from those calculated using the isochrones in Section \ref{subsec:iso} but are of the same order of magnitude, with an average value of $d_\mathrm{av}=4.5\pm4.6\,\kpc$ ($\mu=13.2$). Two stars (C2222241-094912 and C2242408-024953) have very large distances but the errors are also large. If we exclude the largest of these but retain the other for consistency with the isochrone distance average, the mean distance reduces to $d_\mathrm{av}=3.6\pm3.4\,\kpc$ ($\mu=12.8$), which is very similar to the value found for the isochrones.

\subsection{Comparison with other distances}
\label{sec:other_distances}
In Table \ref{tab3} we also list distances derived in \citet{Breddels2010} ($d_\mathrm{B}$), \citet{Zwitter2010} ($d_\mathrm{Z}$) and \citet{Burnett2010} ($d_\mathrm{BB}$), where these distances are all derived from RAVE stellar parameters, employing various methodology. Comparing the above $d_\mathrm{I}$ and $d_\mathrm{R}$ to the distances calculated in \citet{Zwitter2010}, for which 11 of the 15 Aquarius stars have an entry, we find that the isochrone distances agree better with a $\mu\pm\sigma$ for the difference $(d_\mathrm{Z} - d_\mathrm{I})/d_\mathrm{Z}$ of  $23\%\pm20\%$, while for $(d_\mathrm{Z} - d_\mathrm{R})/d_\mathrm{Z}$ we have $-10\%\pm100\%$. This is somewhat unsurprising given that the Zwitter distances and the isochrone distances are both based on RAVE stellar parameters. Interestingly, however, for 6 stars that have distances calculated by the method of \citet{Burnett2010}, the RPM distances fare better: $(d_\mathrm{BB} - d_\mathrm{I})/d_\mathrm{BB}$ gives  $-100\%\pm170\%$ while $(d_\mathrm{BB} - d_\mathrm{R})/d_\mathrm{BB}$ yields  $-30\%\pm70\%$. For the 6 \citet{Breddels2010} entires we have much larger discrepancies of $(d_\mathrm{B} - d_\mathrm{I})/d_\mathrm{B}$ of $-200\%\pm40\%$ and $(d_\mathrm{B} - d_\mathrm{R})/d_\mathrm{B}$ of $330\%\pm60\%$. Clearly, all these discrepancies imply that the individual distances listed in Table \ref{tab3} have large uncertainties. In general, however, the RPM distances give more consistent kinematics than the isochrone distances as we will see below in Section \ref{sec:Pos}. 

\section{Possibly related substructure}
\label{sec:Pos}
In this section we seek connections between the Aquarius stream and other known kinematic and spatial substructures nearby in the Galaxy. We start with the spatially detected substructures before returning to kinematically detected solar neighbourhood features. 

\subsection{Large stellar streams and features}
\label{subsec:large}
The nearest companion of the Milky Way, the Sgr dSph \citep{Ibata1994}, has shed significant debris on its polar orbit around our Galaxy. The all-sky mapping of this debris by \citet{Majewski2003} using 2MASS M-giants clearly showed the plane of the Sgr debris. Further studies such as those by \citet{Newberg2003, Martinez2004, Belokurov2006} have revealed further branches and details within the debris wraps. Recently, \citet{Yanny2009a} used M and K giants selected from SDSS and SEGUE data \citet{Yanny2009b} to provide additional observational constraints on the stream. 

\begin{figure*}[!tb]
\epsscale{1.0}\plotone{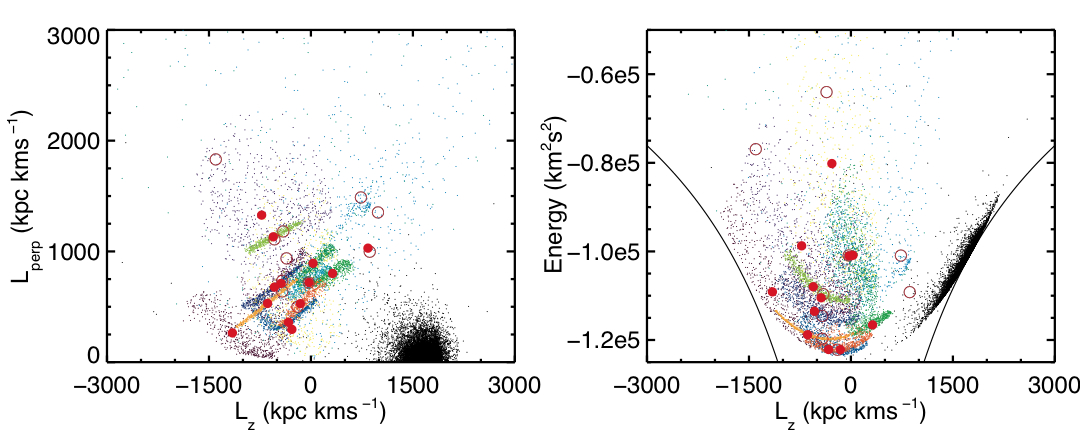}
\caption{(a) $L_Z$-$L_\mathrm{perp}$ and (b) $L_Z$-Energy (Lindblad) plane for the Aquarius stars with the solid red points calculated using $d_\mathrm{R}$ and the open red points $d_\mathrm{I}$. In the background the mostly thin-disk GCS stars are shown as black points. The uncertainty in the solid red points are shown via the clouds of small colored dots, which give the $1\sigma$ spread for the MC simulation. Each color shows the spread for a single solid red point. A high degree of co-variance in the large errors is evident, though the retrograde nature of the Aquarius stream stars is clear. The solid curves in the Lindblad diagram represent the circular orbit loci for this potential. 
\label{f6}}
\end{figure*}

The Aquarius stars fall fairly close to the orbital plane of the Sgr dwarf. Also, the isochrone fit from Section \ref{sec:Pop} is consistent with a population of $10\,\gyr$, $\mh=-1$. \citet{Layden2000} obtain CMDs of Sgr field populations, finding a dominant old and intermediate age population of 11Gyr, $\mh=-1.3$ and 5 Gyr, $\mh=-0.7$. \citet{Giuffrida2010} find a range of populations in the periphery of Sagittarius, with $\mh=-2.34$ to -0.6, while the dominant population has a similar metallicity to 47 Tuc with $\mh=-0.6$. Given the errors, the isochrone fit for the Aquarius stream is consistent with the Sgr dwarf. We thus investigated a possible link between the Aquarius stream and the Sagittarius dwarf debris. The details of this investigation are given in Appendix A. 

The overall result is that the Aquarius stream's kinematics \textit{do not} match those of the Sagittarius dwarf debris, calculated using a variety of potential models (oblate, spheroid, prolate, triaxial) from Law (2005, 2009). The oblate model shows a potential match for a small section of nearby debris when considering the line-of-sight velocity in the Galactic rest-frame, $\vgal$, alone. However, the full kinematics of 
$\VPHI,\ \VR,\ \VZ$ displays that the kinematics of the Aquarius stream and this nearby section are actually quite different\footnote{We use the \citet{Dehnen1998} values for the solar peculiar velocity of $(U,\ V,\ W) = 10,\ 5,\ 7\,\kms$ with respect to the LSR, which we set at a rotation velocity of $220\,\kms$.}. The possible connection is further ruled out by the fact that the oblate halo potential model does not compare well with other observational data for Sgr dwarf debris. 

Since the Aquarius stream lies in the southern part of the RAVE data, it could not be discovered in the main, northern SDSS survey. Thus the stream is far removed from the SDSS-discovered substructures, including the Canis Major overdensity at $(l,\ b)=(240^\circ,\ -8^\circ)$ \citep{Martinez2005} and the Virgo overdensity at $(l,\ b)=(300^\circ,\ +60^\circ)$ \citep{Juric2008}. Further, the stream is located between the southern SEGUE SDSS stripes so it unsurprising that this has not been detected in this survey. The stream's Galactic latitude of $b=-60^\circ$ rules out a relation to the more planar Monoceros stream ($b<40^\circ$) \citep{Penarrubia2005}. Its velocities and latitude are also inconsistent with the thick disk asymmetries detected by \citet{Parker2003, Parker2004}.

The Hercules-Aquila cloud, again detected using SDSS photometry, is located at $l=40^\circ$ and extends above and below the plane by $50^\circ$ \citet{Belokurov2007}. The velocities of the $b>0^\circ$ segment are $\vgal = +180\,\kms$ and the structure ranges over heliocentric distances of $d=10 - 20\,\kpc$. The Hercules-Aquila cloud is near the Aquarius stream on the sky. However, despite the lack of velocity data below the plane, it can be clearly seen that the two entities are separate: the centering in (l, b) for the two are shifted from each other and their distance ranges are clearly incompatible. Additionally, in Section \ref{subsec:sat} we trace the orbit of a simple model for the Aquarius stream and the resulting region of phase-space that the debris inhabits does not overlap with the Hercules-Aquila cloud in ($l,\ b,\ \vgal$). 

\begin{table*}[!bht]
\begin{footnotesize}
\begin{tabular}{llllllllll}
\hline
&$V_R$&$\sigma(V_R)$&$V_\phi$&$\sigma(V_\phi)$&$V_z$&$\sigma(V_z)$&$L_z$&$L_\mathrm{perp}$&Energy\\
&$\kms$&$\kms$&$\kms$&$\kms$&$\kms$&$\kms$&$\kpc\,\kms$&$\kpc\,\kms$&$\energyunits$\\
\hline
A&-50&220&-50&110&110&80&-360$\pm$710&1000$\pm$400&$-8\pm6 \times10^4$\\
B&0&170&-50&70&90&60&-330$\pm$500&710$\pm$320&$-9\pm3 \times10^4$\\
C&0&40&-75&5&60&30&-590$\pm$30&480$\pm$100&$-9.3\pm0.6 \times10^4$\\
\hline
\hline
&$r_\mathrm{peri}$&$r_\mathrm{apo}$&$z_\mathrm{max}$&$e$\\
&$\kpc$&$\kpc$&$\kpc$\\
\hline
A&1.5&9&6&0.8\\
B&1.8&9&5&0.7\\
C&1.8&9&4&0.7\\
\hline
\end{tabular}
\caption{Orbital properties of the Aquarius stream using (A) isochrone distances, (B) RPM distances and (C) the satellite model in Section \ref{subsec:sat}, selected between $-70<b<-50$, $d<5$ kpc, $t_s=-700\,\myr$.}
\label{tab4}
\end{footnotesize}
\end{table*}

\subsection{Solar neighbourhood streams}
\label{subsec:solar}
We have calculated orbits for candidate stars in the potential of Helmi et al (2006), which has contributions from a disk, bulge, and dark halo. Table \ref{tab4} gives averages for various quantities derived from these orbits as well as the median quantities for the overall kinematics, using both sets of distances. Note that we chose the median as it gives more consistent results, and for this reason we also excluded the two most distant stars with $d> 9$ kpc as their kinematics differed greatly from the others. Also, the values for the pericentre and apocentre only include non-radial orbits. 

Figure \ref{f6} shows the $L_z$-$L_\mathrm{perp}$ and  $L_z$-Energy (Lindblad) planes for orbits based on both distance estimates, where to aid comparison to other studies we use here energies as calculated in \citet{Dinescu1999}. Note that the scatter of the isochrone distance results is large so the majority of these points lie off the plot, as do some of the RPM distance results. We plot for reference stars in the Geneva Copenhagen survey \citep{Nordstrom2004}, which is comprised mainly of thin and some thick disk stars. The circular orbit loci for this potential are also shown in the Lindblad diagram. To show the typical error covariance we also ran a Monte Carlo (MC) simulation for each star (with the RPM distances). From the errors in distances, proper motion and radial velocity we generated a sample of 1000 points representative of each distribution, which were then propagated through into the momenta and energy. Following \citet{Wylie2010}, in Figure \ref{f6} we plot the resulting distributions within $1\sigma$ of each of the average values. These demonstrate the large non-Gaussian errors and covariances in $L_z,\ L_\mathrm{perp}$, Energy: the Aquarius stream could not be found initially in these planes. Indeed, all three values are quite ill-constrained with the current uncertainties in the stellar distances. 

Nevertheless, we can at least say that the stars are retrograde and that they are away from the notable halo feature of \citet{Helmi1999} at ($L_z,\ L_\mathrm{perp})=(1000,\ 2000)\,\kpc\,\kms$. The Aquarius stream is also not near the prograde and retrograde features of \citet{Kepley2007} at ($L_z,\ L_\mathrm{perp})\sim(2500,\ 500)\,\kpc\,\kms$ and ($L_z,\ L_\mathrm{perp})\sim(-2500,\ 1700)\,\kpc\,\kms$. Also, with values of $(V_\mathrm{az},$ $V_{\Delta E},$ $\nu) = (125\,\kms,$ $205\,\kms,$ $-15^\circ)$\footnote{\citet{Klement2009} define $V_\mathrm{az}=\sqrt{(V+V_\mathrm{LSR})^2+W^2}$, $V_{\Delta E}=\sqrt{U^2+2(V+V_\mathrm{LSR})^2}$ and $\nu=\arctan((V+V_\mathrm{LSR})/{W})$} it is not one of the newly detected solar neighbourhood streams listed in Table 2 of \citet{Klement2009}. Another solar neighbourhood stream is the Kapteyn group, which \citet{Wylie2010} suggests is stripped from $\omega$ Centauri (also see \citet{Meza2005}), which in turn is thought to be the surviving nucleus of an ancient dwarf galaxy \citep{Bekki2003}. \citet{Wylie2010} employ the same potential as we do here, finding $L_z=-413\,\kpc\,\kms,\ E=-1.3\times10^5\,\energyunits$ for $\omega$ Cen and $ L_z\sim-200\,\kpc\,\kms,\ E\sim-9.5\times10^4\,\energyunits$ for the Kapteyn group. The Aquarius stream is somewhat similar to the distribution in $L_z$-Energy for the Kapteyn group/$\omega$ Cen. However, in Section \ref{subsec:progenitor} we will see that our model for the stream rules against an association. Another significant halo clumping found by \citet{Majewski1996} towards the north Galactic pole has a retrograde orbit with $\VPHI\sim-55\,\kms$, which is consistent with that of Aquarius. The mean $|z|$ for this moving group of $|z_\mathrm{av}|=4.5\,\kpc$ is rather high when compared to ($|z_\mathrm{median}|,\,|z_\mathrm{av}|)\sim(2,\, 3.5)\,\kpc$ for Aquarius, as well as the $z_\mathrm{max}$ values in Table \ref{tab4}. Moreover, our model for the Aquarius stream in Section \ref{subsec:progenitor} does not overlap with the north Galactic pole, again ruling against an association. 

\section{Nature of the stream}
\label{sec:Nat}
The distances calculated for Aquarius stream stars place it fairly close to the sun. If they are of the correct order of magnitude then we could possibly expect additional stream members in other areas of the sky. However, our exploration of the two bounding latitude ranges in Section \ref{subsec:Loc} yielded no striking overdensities that we can immediately associate with the Aquarius stream. Two other regions had overdensities detected for $J>10.3$, though they are not as conspicuous as Aquarius: the region around $-50^\circ<b<-30^\circ,\ 330^\circ<l<345^\circ$, $190<\RV<270\,\kms$ (Region A) and the region $-70^\circ<b<-50^\circ,\ 280^\circ<l<315^\circ$, $40<\RV<130\,\kms$ (Region B). To establish if these additional areas could be associated with the Aquarius stream, and if not, how the stream's localization arises, we created a simple model of a satellite dissolving in the potential of the Galaxy. 

\subsection{Model satellite disruption}
\label{subsec:sat}
To generate a simple satellite dissolution, we first chose one of the average, stable orbits -- that using $d_\mathrm{R}$ for the star C2322499-135351 -- and integrated the orbit back in time. Centering on the orbital positions at various times in the past, $t_s$, we generated $10^4$ test particles from a Gaussian sphere with core radius and internal velocity dispersion $r_c=300$ pc, $\sigma_V=10\,\kms$. Neglecting self-gravity we then integrated the orbits of the satellite forward in time until the present day. This approximate approach suffices as our aim here is illustrative rather than finding the definitive orbit for the Aquarius stream. 

\begin{figure*}[!tb]
\epsscale{1.0}\plotone{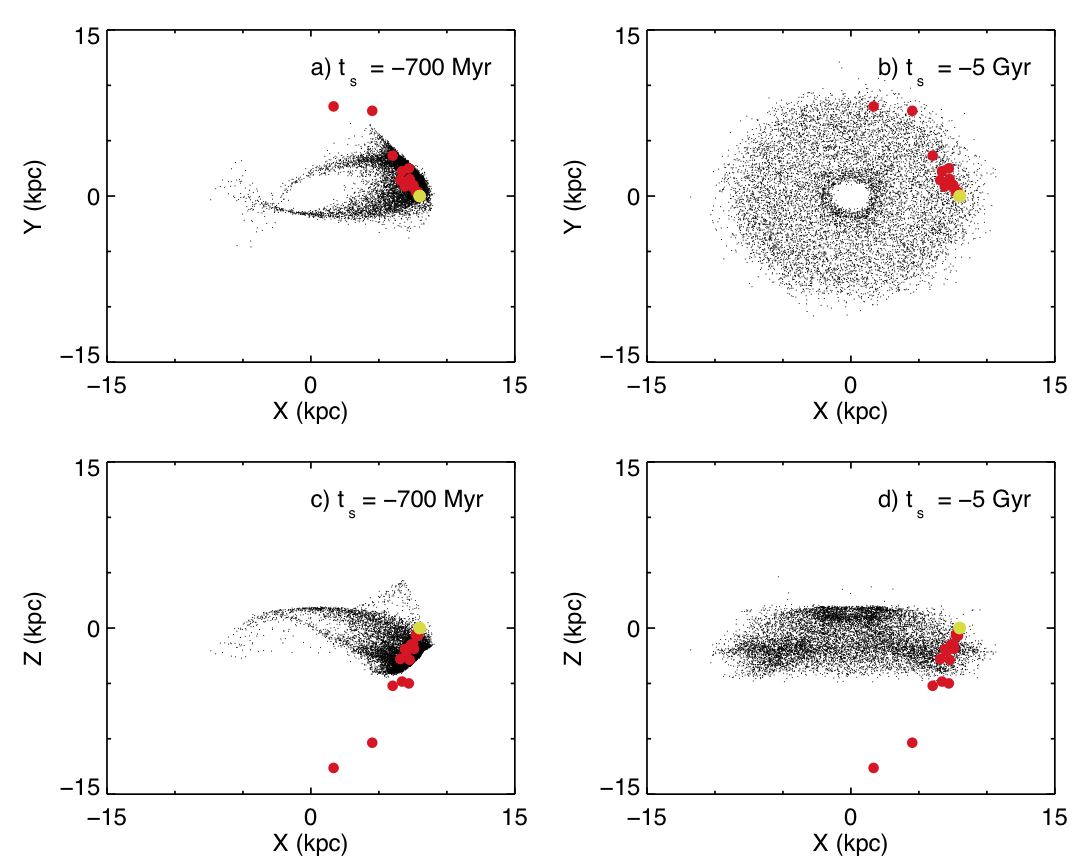}
\caption{$XY$ {(\textit{top})} and $XZ$ {(\textit{bottom})} planes of a simple satellite model of the Aquarius stream at the present day with two different starting times: 
$t _s=-700\,\myr$ {(a, c: \textit{left})} and $t_s=-5\,\gyr$ {(b, d: \textit{right})}. The position of the sun is marked as a yellow point and the Aquarius stars (using $d_\mathrm{R}$) as red points.
\label{f7}}
\end{figure*}

Figure \ref{f7} shows the distribution of particles for two different starting times: one starting two orbital periods ago, $t_s=-700\,\myr$, and another starting at $t_s=-5\,\gyr$. These were selected to illustrate two different extremes, one in which the stream at the present day has yet to be significantly phase mixed and the other when it is completely phase mixed. In both scenarios the Aquarius stream stars sample a volume near the apocenter of the orbits. The furthermost stars have $Z$ values that are substantially larger than $Z_\textrm{max}=4$ kpc for the dissolving satellite, a discrepancy that may be resolved either by a more detailed model or by more accurate distances. In general, however the majority of Aquarius (and RAVE) stars are within a few kpc of the sun, and only a small portion of the total volume traced by the orbits falls within this sample volume. 

\begin{figure*}[!tb]
\epsscale{1.0}\plotone{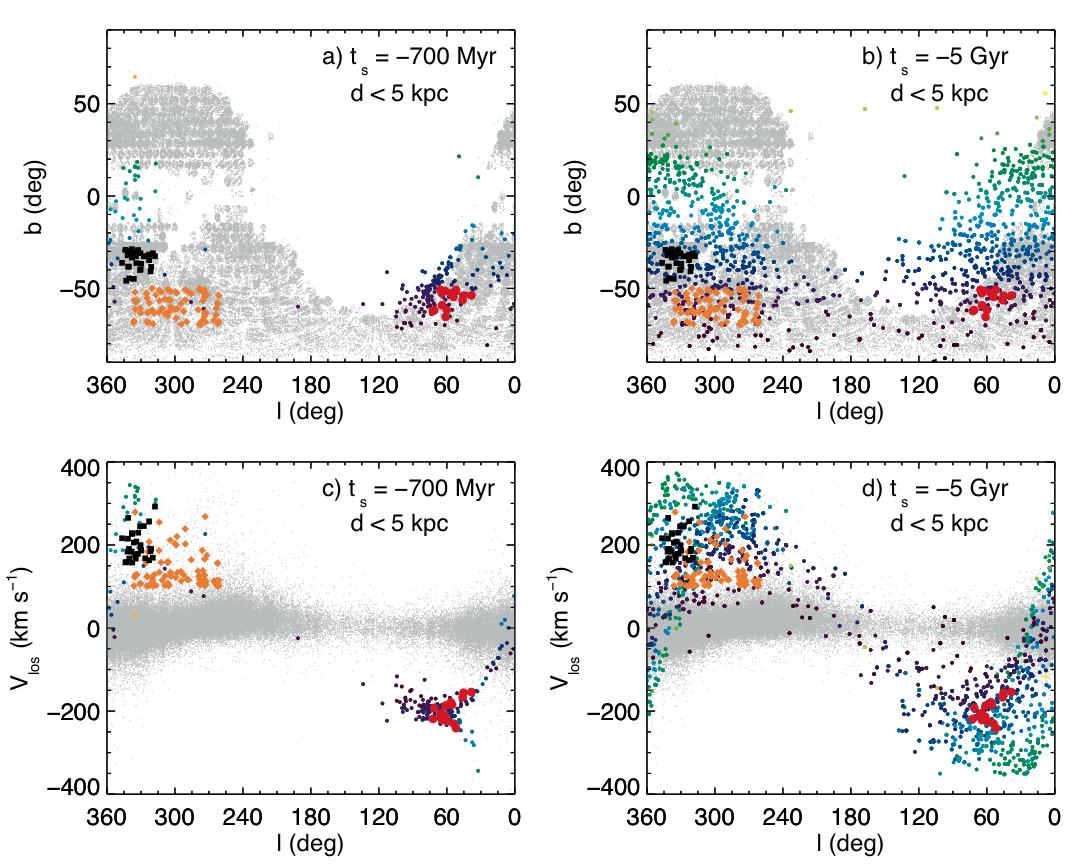}
\caption{The $l$-$b$ (\textit{top}) and $l$-$\vrad$ (\textit{bottom}) planes for the present-day satellite particles within $d<5$ kpc for simulations with two starting times, $t_s =-700\,\myr$ (a, c: \textit{left}) and $t_s=-5\,\gyr$ (b, d: \textit{right}). The test particles are the small colored points with the color denoting their Galactic latitude; those at $b=+20^\circ$ are green, progressing to light blue at $b=-10^\circ$,  dark blue at $b=-40^\circ$ and violet at $b=-70^\circ$. The Aquarius stream stars (large red points) are superimposed as are the two other regions with possible overdensities: $-50^\circ<b<-30^\circ,\ 320^\circ<l<350^\circ$, $150<\RV<300\,\kms$ (Region A: black squares) and $-70^\circ<b<-50^\circ,\ 260^\circ<l<340^\circ$, $100<\RV<300\,\kms$ (Region B: small orange diamonds). RAVE stars having $J>10.3$ are plotted in the background (grey points). The Aquarius stream bears a striking resemblance to the left-hand scenario, suggesting the stream is not yet phase mixed. \label{f8}}
\end{figure*}

In Figure \ref{f8} we plot $l$-$b$ and $l$-$\vrad$ planes for test particles within $d<5$ kpc of the sun for the simulations with the two different starting times, where we chose this distance limit as it encompasses $\sim70\%$ of the Aquarius stars. We also plot the Aquarius candidates and the stars that fall within Regions A and B, as well as the background population of RAVE stars. The distribution of particles for the simulation with $t_s=-700\,\myr$ occupy a small region in $l$-$b$-$\vrad$ that does not coincide with the locations of Regions A and B. Furthermore, the distribution in $l$-$\vrad$ remarkably mimics that of the Aquarius stream and a picture emerges of how the stream can be so localized: with the RAVE survey data we miss portions of the stream due to the location of the survey boundary in some regions and in others, because $\vrad$ overlaps with that of the main distribution so the stream is difficult to detect. 

In the phase-mixed ($t_s=-5\,\gyr$) scenario we see that there is overlap between the region in $l$-$b$ plane occupied by the test particles and for Regions A and B, though the $\vrad$ values for Region B agree better with those of the test particles than Region A. However, it would be difficult to associate both Regions A and B with Aquarius without others regions being populated. In particular, we could expect overdensities at $-50^\circ <b<-30^ \circ $, $l>240^\circ$, $\vrad=+200\,\kms$ as well as a population at $-30^ \circ<b<-50^ \circ$, $l<60^\circ$ out to $\vrad=-200\,\kms$. Such overdensities are not observed in the data.

These simple models of a dissolving satellite therefore suggest that the localization of Aquarius is due to further regions of phase-space not yet being populated: the region in ($l,\ b,\ \vrad$) space occupied by the $t_s=-700\,\myr$ simulation is more consistent with the observed population than that of the $t_s=- 5\,\gyr$ simulation. This suggests that the stream is most likely dynamically young, resulting from a recent disruption of a progenitor, and has not yet undergone phase mixing. Indeed, our simple model of a recently disrupted satellite is very successful in reproducing the main features of the Aquarius stream. In Table \ref{tab4} we therefore list the parameters found for test particles from this model in the Aquarius stream's latitude range. These are quite similar to those found using the $d_\mathrm{I}$ and $d_\mathrm{R}$. The results also corroborate our observation from the RPMD in Section \ref{sec:RPM} that the stream stars have a constant transverse velocity, a fact which was used in in Section \ref{sec:rpmdist} to derive $d_\mathrm{R}$. Test particles for the $t_s=-700\,\myr$ simulation within the Aquarius stream latitude range and distance range, i.e., $-70^\circ<b<-50^\circ$ and $d<5$ kpc have $v_t=250\pm33\,\kms$. This agrees with the value of $v_t=250\pm100$ found from the RPMD.

From Figure \ref{f8} we see that the recent-disruption model suggests that at $-10^ \circ<b<+30^ \circ,\ 330^ \circ<l$ we can expect a smaller population of stars associated with the Aquarius stream out to $\vrad=+350\,\kms$. This begins to overlap at $+20 ^ \circ<b$ with the RAVE survey area, though we do not detect such a population in the data. Future releases of RAVE data with more observations in this area, combined with more careful modelling of the stream, will enable a better understanding of whether this area is indeed populated by Aquarius stream stars.

\subsection{Progenitor of Aquarius}
\label{subsec:progenitor}
The above scenario of a dynamically young stream is not inconsistent with an age of $10\,\gyr$ for the Aquarius candidates, as estimated from the isochrone fit in Section \ref{subsec:mdfcmd}: the stream can be seen as a remnant of an old satellite that has been recently disrupted. As to the nature of the progenitor of the Aquarius stream, it could either be a dwarf galaxy or a globular cluster. The survival of this progenitor would depend on its concentration: it could either have been tidally stripped or have undergone complete disruption. 

To search for possible globular clusters that the Aquarius stream could have been tidally stripped from, we performed a search of the \citet{Harris1996} catalog of known globular clusters, selecting those with $-1.5<\mathrm{[Fe/H]}<-0.5$, $1.5<R_\mathrm{GC}<9$ kpc and $Z<4$ kpc, where the latter limits are taken from the model satellite orbit in Table \ref{tab4}. We then compared the distribution of the clusters in $l,\ b, \vrad$ to that of the model satellite stream ($t_s=-700\,\myr$), and found no globular clusters that match the simulation. Also, $\omega$ Cen, with an $l,\ b,\ \vrad$ of $(309^\circ,\ +15^\circ,\ 233\,\kms)$ and $L_z=-413\,\kpc\,\kms$, is not consistent with not-yet-phased-mixed scenario in Figure \ref{f8} and Table \ref{tab4}. With our uncalibrated metallicities it is difficult to compare the MDF to that of $\omega$ Cen. A high-resolution spectroscopic abundance study, such in \citet{Wylie2010}, is required (as well as further modelling of Aquarius) to definitively understand if Aquarius is related to $\omega$ Cen. 

The progenitor of the Aquarius stream therefore is currently unknown. 

\section{Conclusion}
In this paper we report the detection of a new halo stream found as an overdensity of stars with large heliocentric radial velocities in the RAVE data-set. The detection is enabled by RAVE's selection criteria creating no kinematic biases. The fifteen member stars detected have $\RV=-199\pm27\,\kms$ and lie between $-70^\circ< b <-50^\circ$, $30^\circ<l<75^\circ, J>10.3$ in the constellation of Aquarius. We established the statistical significance of the stream by comparing the RAVE data, in the Galactic latitude range $-70^\circ<b<-50^\circ$, to equivalent mock samples of stars created using the Besan\c con Galaxy model and the code Galaxia. For different cell sizes, $\Delta l \times \Delta \vrad$, we compared the number of stars in the data and models, finding that for the majority of cell sizes the region around the Aquarius stream exhibited a $4\sigma$ overdensity in the data, irrespective of the dust modelling. Searching for additional overdensities in neighboring latitude regions yields no structures of the same level of significance (other than the LMC), though two regions are identified as being marginally overdense.  

For most of the Aquarius stars RAVE stellar parameter estimates are also available. The member stars are metal-poor with $\mh=-1\pm 0.4$ and we derive a preliminary isochrone fit in the $\teff$-$\log g$ plane with an population age of $10\,\gyr$. Both the $\vrad$ and metallicity are consistent with the group being within the stellar halo. We further use a Reduced Proper Motion Diagram to derive the transverse velocity for the stream, finding $v_T=250\pm100\,\kms$ for the group. This again places it within the Galaxy's halo. We use the isochrone fits and the RPMD to provide distance estimates to the stars, where we prefer the latter as they give more consistent kinematics.

We investigated the relation of the stream to known substructures. We first discussed the probability of the stream being with debris from the Sagittarius dwarf. This is a priori plausible because the stream does not fall far from the orbital plane of the Sgr dwarf and the stream's metallicity is consistent with that of the dwarf. A comparison to the models of Sagittarius dwarf debris from \citet{Law2005} and \citet{Law2009}, shows that although the majority of models do not yield a good fit, a certain selection of nearby stars in the oblate model provides a reasonable fit in the $\Lambda_\sun$-$\vgal$ plane. This is most likely just coincidental however: the distributions in both distance and $\VPHI$ are clearly inconsistent with those of the Sgr stream. Also, the oblate model is the least favoured of all the models when compared to the most recent data for the Sagittarius stream. We thus conclude that the Aquarius stream is most likely not associated with the Sagittarius dwarf. A search of other known substructures both in the solar neighbourhood (e.g. Kapteyn group) and in the solar suburb (e.g. Canis Major and Virgo overdensities, Monoceros stream, Hercules-Aquila cloud) yielded no positive identifications. 

Finally, to understand better how the stream is both local and localized on the sky, we performed simple dynamical simulations of a model satellite galaxy dissolving in the Galactic potential. We presented simulations for two time-scales, one where the satellite is dissolving and the other when it is completely phase mixed. We compared the distribution in $l,\ b, \vrad$ space of nearby tracer particles at the present day to that of the Aquarius stream stars plus the two other marginally overdense regions found in the RAVE data. The model in which the progenitor has had time to become phase mixed predicts over-densities in places were the data show none. By contrast, the dissolving, not-yet-phase-mixed scenario was able to account for the localization as well as reproducing the observed structure of the Aquarius stream. We therefore suggest that the stream is dynamically young: the localization could be explained as a recent disruption event of a progenitor whereby the stream has yet to occupy the available phase-space. The progenitor could either be a globular cluster or a dwarf galaxy, which may or may not have survived to the present day. We make no positive identification of with any globular clusters, though there could be a possible link with likely dwarf galaxy remnant, $\omega$ Cen, and the associated Kapteyn group. Follow-up high-resolution abundances would elucidate this possible connection. Further, more sophisticated simulations of Aquarius are required. This will enable a better understanding of this interesting, new halo stream which places hierarchical formation right on our proverbial doorstep. 

\acknowledgments
Funding for RAVE has been provided by: the Anglo-Australian Observatory; the Astrophysical Institute Potsdam; the Australian National University; the Australian Research Council; the French National Research Agency; the German Research foundation; the Istituto Nazionale di Astrofisica at Padova; The Johns Hopkins University; the National Science Foundation of the USA (AST-0908326); W.M. Keck foundation; the Macquarie University; the Netherlands Research School for Astronomy; the Natural Sciences and Engineering Research Council of Canada; the Slovenian Research Agency; the Swiss National Science Foundation; the Science \& Technology Facilities Council of the UK; Opticon; Strasbourg Observatory; and the Universities of Groningen, Heidelberg and Sydney. The European Research Council has provided financial support through ERC-StG 240271(Galactica).

M E K Williams would like to thank David Martinez-Delgado for stimulating discussions and the anonymous referee for their helpful comments. 

{\it Facilities:} \facility{UKST}

\newpage
\appendix

\section{Ruling out Sagittarius dwarf debris}
\label{sec:Sgr}
\begin{figure}[!tb]
\epsscale{0.7}\plotone{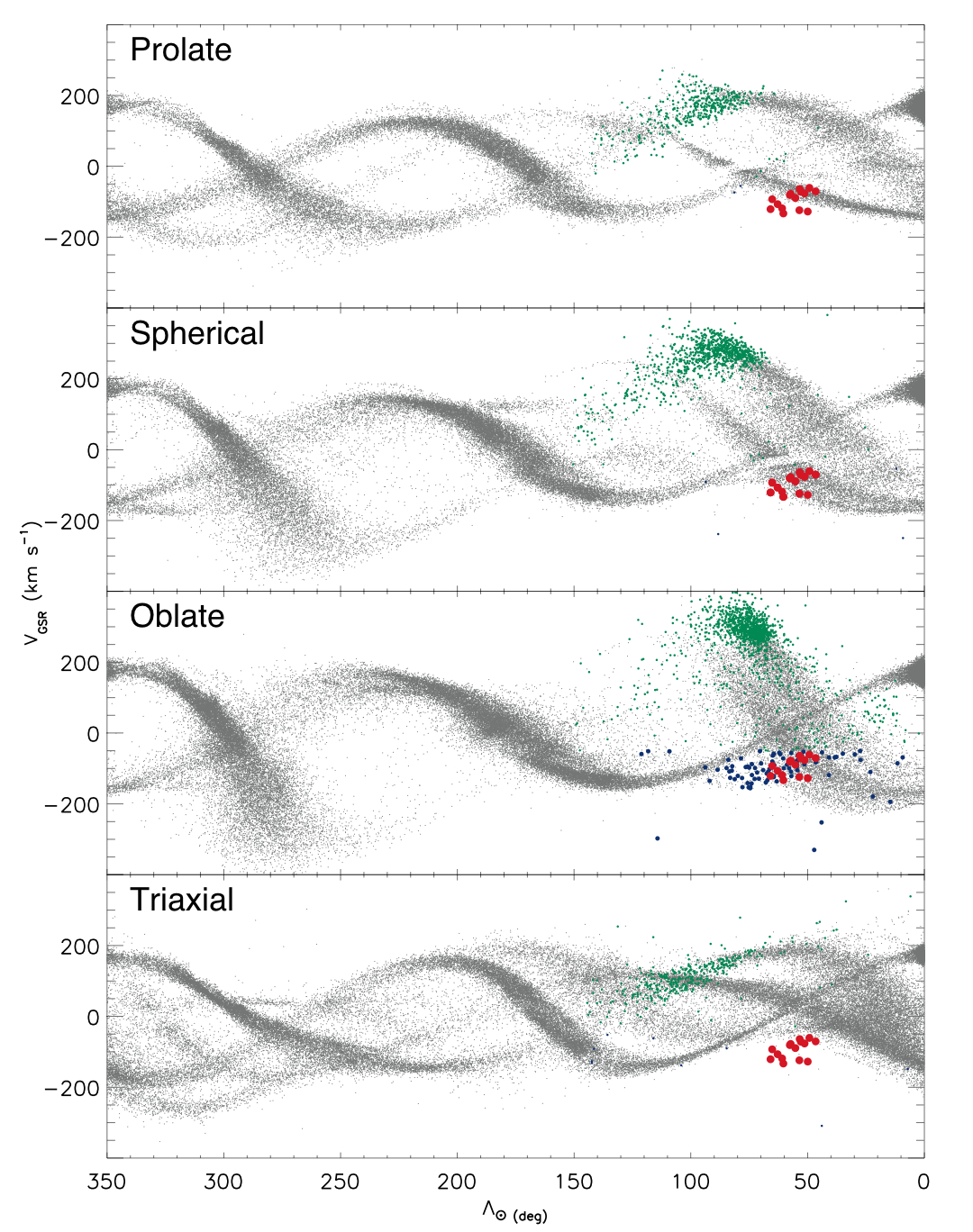}
\caption{Longitude in the Sagittarius orbital plane, $\lamsun$, vs galactocentric radial velocity for the Aquarius stream stars (red points) compared to the Law models. Stars in the models with $d<15\,\kpc,\ b<0^\circ,\ l<0^\circ,\ \vgal>-50\,\kms$ are labelled green while  $d<15\,\kpc,\ b<0^\circ,\ l<0^\circ,\ \vgal<-50\,\kms$ are blue. In this plot the blue points in the oblate model show a possible match to the Aquarius stream stars. \label{f9}}
\end{figure}

We compared the kinematic properties of the Aquarius stream stars to models of the Sagittarius dwarf debris, using the prolate (q=$0.9$), spherical (q=$1.0$) and oblate (q=$1.25$) models of \citet{Law2005} (hereafter L05)\footnote{available from http://www.astro.virginia.edu/~srm4n/Sgr/}. We also used the triaxial model of \citet{Law2009} (hereafter L09) which uses a halo potential with $c/a \approx 0.67,\ b/a\approx0.83$ within $60\,\kpc$\footnote{kindly provided before public release by D. Law}. We follow convention and use the parameter $\lamsun$ (defined in Figure 1 of L05) in our plots, which is the longitude from the Sgr dSph in the plane of its orbit increasing away from the Galactic plane. Figure \ref{f9} plots $\lamsun$ against Galactocentric radial velocity for the Aquarius stream stars and the different Law models. We highlight those stars from the models that have the following properties:

\begin{itemize}
\item distance to sun $< 15\,\kpc$
\item declination $<0^\circ$
\item Galactic latitude $<0^\circ$
\end{itemize}

We are generous with the distance and Galactic latitude criteria to allow for uncertainties in the models (as well as the observational distances). Figure \ref{f9} shows $\vgal$ as a function of $\lamsun$. Those stars that fit the above selection criteria and that have $\vgal > -50\,\kms$ are marked in green while those that have $\vgal < -50\,\kms$ are marked in blue. The reason for this delineation will become evident below.

The velocities show a main feature at $\vgal=+200-250\,\kms$, which corresponds to the leading arm, which the models predict is vertically streaming through the solar neighbourhood. This stream gets stronger going from the prolate towards the oblate models. This large signal is not present in the RAVE data, as the heliocentric radial velocity is of the order of $\RV=-270\,\kms$ or $W=-300\,\kms$. \citet{Seabroke2008} showed that there is no such large asymmetry detectable in the distribution of radial velocities for stars with $l<-45^\circ$. We do see, however, that a faint signal of stars is present for the oblate models with $\vgal=-100\,\kms$, which is associated with an extra wrap of the leading arm passing south of the solar neighbourhood in this model. A value of $\vgal=-50\,\kms$ separates this extra, fainter wrap from the main leading arm component predicted by the oblate model. The triaxial model does not exhibit this feature and indeed the $\vgal=+300\,\kms$ stream is weaker as the leading arm misses the solar neighbourhood.

\begin{figure}[!tb]
\epsscale{1.0}\plotone{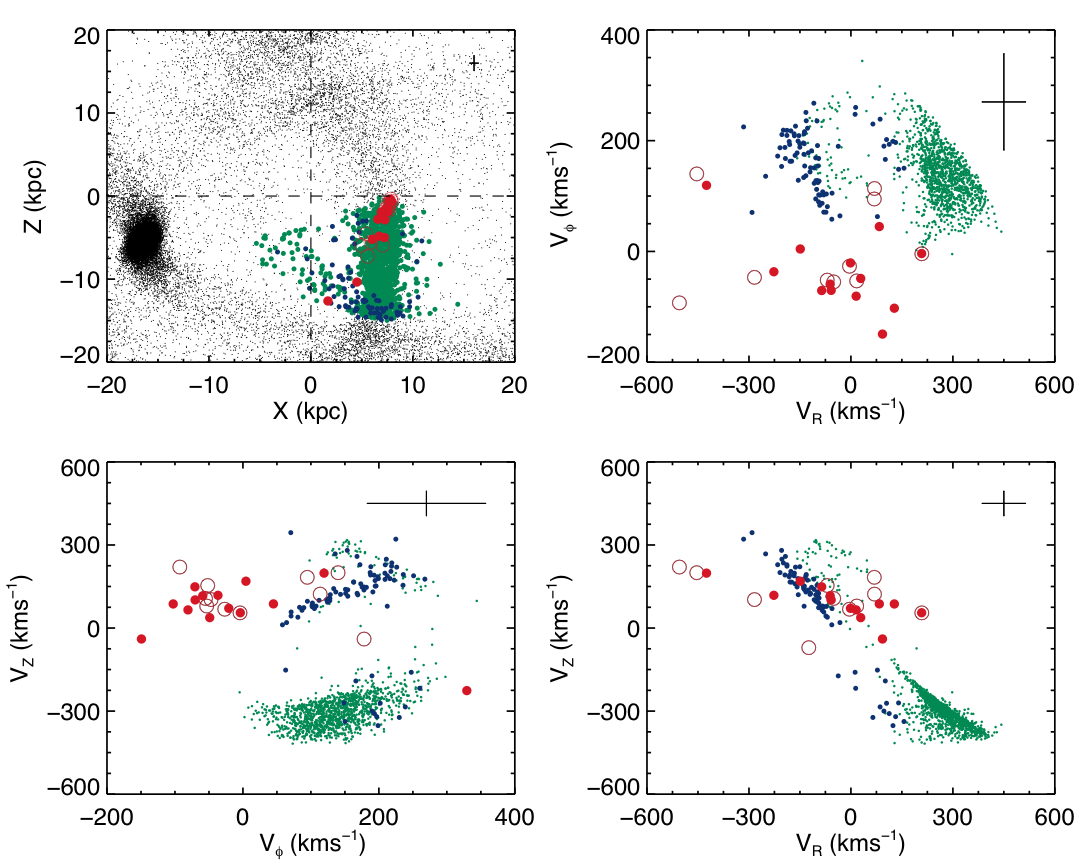}
\caption{$XZ$ plane and $V_\phi$, $V_R$ and $V_Z$ plots showing the Aquarius stream stars and the oblate model of L05. The solid red points show the values for Aquarius stream stars using the RPM distances while the open red points use the isochrone derived distances. Median errors bars for the RPM distance derived values are also shown. The blue and green points are as in Figure \ref{f9}, where we see that with the full 6D phase-space information the possible match with the blue points from the oblate model is ruled out.
\label{f10}}
\end{figure}

We concentrate on the oblate model with its possible curl of the leading arm fitting the Aquarius stream stars, as this is the only possible match. In Figure \ref{f10} we show the $X$-$Z$ plane as well as the $\VR$-$\VPHI$-$\VZ$ planes for the stream stars and the solar neighbourhood stars from the oblate model. The space velocities were calculated using the radial velocities and proper motions in the RAVE catalog as well as the distances derived via the two different methods (isochrone and RPMD). There is some overlap between the velocities from the two different distance derivations but in general we see that the space velocities are affected by the uncertainties in the distances. The RPM distances give a much tighter grouping in velocity for the stars and we take these results to be more indicative of the group's properties, plotting median error bars for these values.

The first thing to note is that the $\vgal<-50\,\kms$ simulation particles do not fit the positions of the Aquarius stream stars in the $X$-$Z$ plane. Even accounting for distance errors the distribution is strikingly different. Rather, the group stars tend to be aligned spatially with the $\vgal>-50\,\kms$ simulation particles. Secondly, while the $\VR$ and $\VZ$ values for the $\vgal<-50\,\kms$ simulation particles are similar to that of the group, the values for $\VPHI$ very much differ from those of the Aquarius stream stars. For while the errors for the stream's $\VPHI$ values are larger than the other velocity components, a significant ($\sim2\sigma$) and systematic shift would be required in this component for the stream and $\vgal<-50\,\kms$ simulation particles to agree. So while there does appear to be some overlap between this faint wrap and the group in a couple of variables, both the spatial distribution and the velocity distribution do not match. This is further borne out by the proper motions: the average value for the Aquarius stream stars is $28\,\masyr$ while for the $\vgal<-50\,\kms$ particles it is $4\,\masyr$.

It is further worth noting that the oblate halo potential model does not compare well to Sagittarius dwarf debris data. As noted by \citet{Fellhauer2006}, the \citet{Belokurov2006} data set traces dynamically old Sgr stream stars around the North Galactic Cap, where the oblate and prolate dark halos give different predictions. These data do not favour the oblate model with \citet{Fellhauer2006} arguing for a spherical dark halo, while L09 favour a triaxial halo. Furthermore, the absence of the Sgr stream near the Sun \citep{Seabroke2008, Newberg2009} is consistent with simulations of the disruption of Sgr in nearly spherical and prolate Galactic potentials. Thus, the only model that has a passing resemblance to the Aquarius stream stars is the least likely of all those presented. On the strength of all the evidence, we conclude that the Aquarius and Sagittarius streams are unrelated.

\end{document}